\documentclass[a4paper,11pt]{article}
\pdfoutput=1 

\usepackage{jheppub} 

\usepackage[T1]{fontenc} 

\title{\boldmath \Large Solution  of the equations of motion for a super non-Abelian sigma model in curved background by the super Poisson-Lie T-duality}

\author[]{Ali Eghbali}

\affiliation[]{Department of Physics, Faculty of Basic Sciences,  Azarbaijan Shahid Madani University,\\ 53714-161, Tabriz, Iran}

\emailAdd{a.eghbali@azaruniv.edu}

\abstract{The equations of motion of a super non-Abelian T-dual sigma model on the  Lie supergroup $(C^1_1+A)$
in the curved background are explicitly solved  by the super Poisson-Lie T-duality.  To find the
solution of the flat model  we use  the transformation of
supercoordinates, transforming the metric into a constant one, which is shown  to be a supercanonical
transformation. Then, using the super
Poisson-Lie T-duality transformations and  the dual decomposition of
elements of Drinfel'd superdouble,  the solution of the
equations of motion for the dual sigma model is obtained. The general form of
the dilaton fields satisfying the vanishing $\beta-$function equations of the
sigma models is found. In this respect, conformal invariance of the sigma models built on  the  Drinfel'd superdouble
$\left((C^1_1+A)~,~I_{(2|2)}\right)$ is guaranteed up to one-loop, at least.}


\begin{document}
\maketitle
\flushbottom

\section{Introduction}
\label{intro}
Non-linear sigma models with target space supermanifolds possess a number of interesting properties.
They are of importance both in superstring theory \cite{Henneaun}  and condensed matter physics \cite{Read}. Perhaps the main motivation
to study them comes from the AdS/CFT  duality.  Furthermore, it has been shown that sigma models on the coset supermanifolds can be used for quantizing
superstring theory with Ramond-Ramond backgrounds. The well-known example is type IIB Green-Schwarz superstring  considered on $AdS_5\times S^5$
background in terms of supercoset formalism \cite{Tseytlin} (see, e.g., \cite{Berkovits}-\cite{Zwiebach}).
In order to have a better understanding of how superstring theory might be quantized, some examples on the Lie supergroups $PSL(n|n)$ were
considered in \cite{Berkovits}, \cite{Bershadsky} and \cite{Boer}. These Lie supergroups, having vanishing Killing form and  Ricci-flat property,  guarantee that the corresponding sigma models are conformally invariant up to one-loop \cite{Kagan}; this is a result of \cite{Bershadsky} in which these models are exactly conformal.

On the other hand, duality transformations have played an important role in  string theory. Target
space duality (T-duality) as a very
important symmetry of string theories - or more generally
two-dimensional sigma models - has been achieved in
revealing the consequences of the isometry symmetry in string theory \cite{Buscher}-\cite{Kiritsis}.
In particular, it  states the equivalence between string theories propagating on two different target
spaces that contain some Abelian isometries \cite{Buscher}.
Unlike the Abelian case \cite{Rocek}, the non-Abelian T-duality transformation is not invertible since
the isometry symmetries of
the original theory are not preserved. So it was not possible to perform the inverse duality transformation to get back to the original model \cite{Ossa}. A solution to this problem was proposed by Klim\v{c}\'{\i}k and \v{S}evera \cite{klse:dna} where it was argued that the two models were dual in the  sense of the so-called Poisson-Lie T-duality.
The Poisson-Lie T-duality  does not require the existence of
isometry in the original target manifold. In this case, the model is not
required to has an isometry symmetry, but there is still an action of a Lie group G on the target
manifold and the Noetherian currents associated with this action are required
to be integrable, that is, they satisfy Maurer-Cartan
equations with group structure of $\tilde G$ (with the same dimension of $G$)  so that $G$ and
$\tilde G$ have Poisson-Lie structure and their Lie algebras are
dual to each other in the sense that they define a bialgebra structure \cite{Drinfel'd}.

The solution of the sigma models in  time-dependent and curved backgrounds is often very complicated. In Ref. \cite{klse:dna} the procedure
for transforming the solutions of a sigma model to those of a dual one has been described. Afterwards, this procedure  was  extended in such a way that
by use of the transformation of group coordinates of the flat model to those for which the metric is constant, a classical solution of the equations of motion for a sigma model in curved background was found in \cite{hla:slnbytduality} (see, also,  \cite{hla1}). Moreover, prior to the procedures done in
\cite{hla:slnbytduality} and  \cite{hla1}, conformally invariant three-dimensional sigma models
on solvable Lie groups, which were Poisson-Lie T-dual or plural to sigma models in the flat
background with the constant dilaton, were investigated in \cite{hlasno:3dsm2} (see, also, \cite{hla2}).

We have recently generalized Piosson-Lie symmetry on manifolds to supermanifolds and have constructed super Poisson-Lie T-dual sigma models
on Lie supergroups \cite{ER}, specially on supermanifolds \cite{ER5}. Furthermore, we have shown that the super Poisson-Lie T-duality relates the WZW models
based on the Lie supergroups to each other \cite{ER8} (see, also, \cite{ER7} ).
In this paper, we are going to solve the equations of motion of a
super non-Abelian T-dual sigma model in curved background by the super Poisson-Lie T-duality. To this end,
we use the fact that  these models in curved background
can be transformed into the flat ones.  By solving partial differential equations that follow  the transformation properties of the connection components
we find the transformation between supergroup coordinates of the model in curved background and Riemannian coordinates (flat supercoordinates)
for which the metric is constant. The model  is built on the Lie supergroup $(C^1_1+A)$. The Lie superalgebra $({\cal C}^1_1+{\cal A})$ \cite{B} of the Lie supergroup $(C^1_1+A)$ is a $(2|2)$-dimensional  Lie superalgebra with two bosonic  and two fermionic generators.
Because of performability of the super Poisson-Lie T-duality transformations, namely finding coordinates of the dual decompositions which depend
on the complexity  of the structure of Drinfel'd superdouble, we have chosen one of the Lie sub-supergroups of the decomposition
of the Drinfel'd superdouble to be Abelian Lie supergroup. In this case,  the super Poisson-Lie T-duality reduces to the super non-Abelian T-duality.
By reducing flat supercoordinates to the solution of the wave equation we obtain solution of the  equations of motion of the original model. We, furthermore, show that the transformation of the flat supercoordinates  is a supercanonical transformation. In the following, by using the super Poisson-Lie T-duality
transformation that follows two possible decomposition of elements of the Drinfel'd superdouble, we get the solution of the equations of motion  for the dual model. The reason why it is very desirable to find the transformation between the supergroup coordinates and the flat supercoordinates  is that in the flat supercoordinates the metric becomes constant. Therefore, the vanishing $\beta$-function equations become very simple. On the other hand, finding solution for conformally invariant sigma models in curved backgrounds is generally a difficult problem. Thus, with regard to the above discussion we easily solve equations for the dilaton field of the flat sigma model  in the flat supercoordinates. Then, to get the general dilaton of the original sigma model which satisfies the vanishing $\beta$-function equations in curved background, we use the flat supercoordinates transformation. Finally, we obtain the general form of the dilaton field satisfying the vanishing $\beta$-function equations for the dual model in terms of the coordinates of the dual Lie supergroup.

The plan of the paper is as follows. In  Section 2, we briefly review  the super
Poisson-Lie symmetry and mutual super Poisson-Lie T-dual sigma models on  Lie supergroups based on our previous papers.  In Section 3, we introduce
the Lie superalgebra $({\cal C}^1_1+{\cal A})$ and the Lie superalgebra of Drinfel'd superdouble
$\left(({\cal C}^1_1+{\cal A}) ~, ~{\cal I}_{(2|2)}\right)$. Moreover, in  Section 3,  the $(2|2)$-dimensional
super non-Abelian T-dual sigma models based on the  Drinfel'd superdouble $\left((C^1_1+A)~,~I_{(2|2)}\right)$
are constructed. Section 4 is devoted to the presentation of a method for finding  flat supercoordinates.  In Section 5  we find the transformation of supercoordinates which makes the metric constant. Also  in Section 5  we prove that the transformation of the flat supercoordinates is indeed a
supercanonical transformation. In Section 6, utilizing the solution of the wave equation and using the super Poisson-Lie T-duality transformations we get
the solution of the equations of motion for the dual sigma model. Finding the general form of
the dilaton fields satisfying the vanishing $\beta-$function equations of the
sigma models is given in Section 7. Finally, we present our conclusions.

\section{A review of the super Poisson-Lie symmetric sigma models on Lie supergroups}

To set our notation, let us briefly review the construction of the super Poisson-Lie T-dual sigma models by means of Drinfel'd superdoubles.
Consider a two-dimensional sigma model action on a supermanifold $M$
as\footnote{ We note that ${|\Upsilon|}$ denotes
the parity of $\Upsilon$ where ${|\Upsilon|}=0$ for the bosonic coordinates and ${|\Upsilon|}=1$ for the fermionic coordinates; here and in the following we use the notation \cite{D}, e.g.,  $(-1)^{|\Upsilon|}:=(-1)^{\Upsilon}$.}
{\small \begin{equation}\label{A1}
S\;=\;\frac{1}{2}\int_{{\Sigma}} d\sigma^{+} d\sigma^{-}\;(-1)^{^{|\Upsilon|}} \partial_{+}\Phi
^{^{\Upsilon}}\;  {
{\cal{E}}}\hspace{-0.5mm}_{_{
\Upsilon \Lambda}}(\Phi)\;\partial_{-}\Phi^{^{\Lambda}},
\end{equation}}
where ${\Sigma}$ is the worldsheet and $\partial_{\pm}$ are the derivatives with respect to the
standard lightcone variables $\sigma^{\pm} :=
\frac{1}{2}(\tau \pm \sigma)$ on the worldsheet, and ${{\cal {\cal
{E}}}}_{_{\Upsilon\Lambda}}\;=\;G_{_{\Upsilon\Lambda}}
+B_{_{\Upsilon\Lambda}}$.  \footnote{Note that $G_{_{\Upsilon\Lambda}}$
and $B_{_{\Upsilon\Lambda}}$ are  the components of the supersymmetric metric $G$ and the
anti-supersymmetric tensor field $B$, respectively,
$$
G_{_{\Upsilon\Lambda}}\;=\;(-1)^{^{\Upsilon \Lambda}}\;G_{_{\Lambda\Upsilon}},~~~~~~~~~~~B_{_{\Upsilon\Lambda}}\;=\;-(-1)^{^{\Upsilon \Lambda}}\;B_{_{\Lambda\Upsilon}}.
$$
We will assume that the metric ${_{_{\Upsilon}}G_{_{\Lambda}}}$ is
superinvertible and its superinverse is shown by
$G^{^{\Upsilon\Lambda}}$.} The coordinates $\Phi{^{^ \Upsilon}}$
include the bosonic  and the fermionic
coordinates, and the label $\Upsilon$  runs over $\mu=0,\cdots,d_B-1$ and  $\alpha=1,\cdots,d_F$ where $d_B$ and $d_F$
indicate the dimension of the bosonic coordinates
and the fermionic coordinates, respectively. Thus, the
superdimension of the supermanifold is written as $(d_B|d_F)$. We define, respectively, the Christoffel symbols (the
components of the connection) and the torsion  as\footnote{\emph{Notation:} If $f$ be a differentiable function on ${\mathbf{R}}_c^m \times {\mathbf{R}}_a^n$ (${\mathbf{R}}_c^m$ are subset of all real numbers with dimension $m$ while ${\mathbf{R}}_a^n$ are subset of all
odd Grassmann variables with dimension $n$), then,  the relation between the left partial differentiation
and right one is given by
$$
{_{_{\Upsilon}}{\overrightarrow \partial}} f := \frac{\overrightarrow{\partial}f}{{\partial} \Phi^{^{\Upsilon}}}  \;=\; (-1)^{\Upsilon(|f|+1)}\;  \frac{f \overleftarrow{\partial}}{{\partial} \Phi^{^{\Upsilon}}},
$$
where $|f|$ indicates the grading of $f$ \cite{D}. }
\begin{eqnarray}\label{A2}
{\Gamma}^{^{\Upsilon}}_{_{\;~\Lambda \Xi}} &=& \frac{(-1)^{^{\Delta}}}{2}G^{^{\Upsilon \Delta}}
\Big[\frac{G_{_{\Delta \Lambda}} \overleftarrow{\partial}}{\partial
\Phi^{^{\Xi}}}+(-1)^{^{\Lambda  \Xi}}~\frac{G_{_{\Delta \Xi}} \overleftarrow{\partial}}{\partial
\Phi^{^{\Lambda}}} -\
(-1)^{^{\Delta(\Lambda+\Xi)}} ~\frac{G_{_{\Lambda \Xi}} \overleftarrow{\partial}}{\partial
\Phi^{^{\Delta}}}\Big],\\
{H}^{^{\Lambda}}_{_{\;~\Upsilon \Xi}} &=&
\frac{(-1)^{^{\Delta}}}{2}G^{^{\Lambda \Delta}} \Big[\frac{B_{_{\Delta
\Upsilon}}\overleftarrow{\partial}}{\partial
\Phi^{^{\Xi}}}+(-1)^{^{\Xi(\Upsilon+\Delta)}}~\frac{B_{_{\Xi
\Delta}}\overleftarrow{\partial}}{\partial
\Phi^{^{\Upsilon}}} +\ (-1)^{^{\Delta(\Upsilon+\Xi)}}~\frac{B_{_{\Upsilon
\Xi}}\overleftarrow{\partial}}{\partial \Phi^{^{\Delta}}}\Big].\label{A3}
\end{eqnarray}
Thus, one  gets the equations of motion for the action (\ref{A1}) as
\begin{equation}\label{A4}
{\partial}_+ {\partial}_-\Phi^{^{\Upsilon}} +
({\Gamma}^{^{\Upsilon}}_{_{\;\;\Lambda\Xi}}-{H}^{^{\Upsilon}}_{_{\;\;\Lambda\Xi}})\;{\partial}_-\Phi^{^{\Xi}}
{\partial}_+\Phi^{^{\Lambda}}\;=\;0.
\end{equation}
If the Noether's current one-forms corresponding
to the right action of the Lie supergroup $G$ on the target space $M$
are not closed and  satisfy  the Maurer-Cartan equation \cite{D} on the extremal surfaces, we say that the
sigma model (\ref{A1}) has the \emph{super Poisson-Lie symmetry} with respect to
the Lie supergroup $\tilde G$ (the dual Lie supergroup to ${ G}$)
\cite{ER}. It is a condition that is given by the following relation \cite{ER} \footnote{The superscript ``$st$'' means supertransposition of the matrix \cite{D}.}
\begin{equation}\label{A5}
{\cal L}_{_{V_i}}({\cal E}_{_{{\Upsilon\Lambda }}})
=(-1)^{i(\Upsilon+k)}\;
 {\cal E}_{_{{\Upsilon\Xi
}}} {(V^{st})^{^{\Xi}}}_{k}\;(\tilde
{\cal{Y}}_i)^{kj}\;{{_jV}}^{^{\Omega}}\;{\hspace{-0.5mm}_{_{\Omega}}
{ {\cal{E}}}}\hspace{-0.5mm}_{_{ \Lambda}},
\end{equation}
where ${\cal L}_{_{V_i}}$ stands for the Lie derivative corresponding to the left invariant supervector
fields ${V_i}$ (defined with left derivative) \cite{ER} and
$(\tilde {\cal{Y}}_i)^{jk} = -{{\tilde f}_{\;\;\;\;i}}^{jk}$
are the adjoint representations of Lie superalgebra $\tilde {\cal G}$ (the dual Lie superalgebra to ${\cal G}$). When ${\cal L}_{_{V_i}}({\cal E}_{_{{\Upsilon\Lambda }}})=0$,
the super Poisson-Lie T-duality is replaced by the super non-Abelian T-duality, i.e.,
the Lie supergroup $G$ is a
isometry supergroup of $M$ and we have the conserved currents.

Let us consider the dualizable sigma models defined on the Lie supergroup space (the generalization to the case when a Lie supergroup
$G$ acts freely on the target space). The super Poisson-Lie dualizable sigma models can be formulated on a Drinfel'd superdouble $D\equiv (G , \tilde G)$ \cite{ER4}, a Lie supergroup whose ${\cal D}$ admits a decomposition
${\cal D}= {\cal G} \oplus {\tilde {\cal G}}$ into a pair of sub-superalgebras maximally isotropic with respect to a supersymmetric and
ad-invariant non-degenerate  bilinear form $<.~ , ~.>$ which is defined by the brackets
\begin{equation}\label{A10}
0~=~< X_i , X_j >\; =\; < {\tilde X}^i , {\tilde X}^j
>,~~~{\delta}^i_{~~j}~=~< {\tilde X}^i  , X_j >\;=\;(-1)^{ij}< X_j , {\tilde X}^i >.
\end{equation}
Such decomposition
is called Manin supertriple $({\cal D}, {\cal G}, {\tilde {\cal G}})$.
The generators of ${\cal G}$ and ${\tilde {\cal G}}$  are, respectively, denoted to
$X_i$ and $\tilde X^i$ and satisfy the (anti)commutation relations \cite{ER1}
$$
[X_i , X_j] = {f^k}_{ij} X_k,\hspace{20mm} [\tilde{X}^i ,\tilde{ X}^j] ={{\tilde{f}}^{ij}}_{\; \; \: k} {\tilde{X}^k},
$$
\begin{equation}\label{A5.1}
[X_i , \tilde{X}^j] =(-1)^j{\tilde{f}^{jk}}_{\; \; \; \:i} X_k
+(-1)^i {f^j}_{ki} \tilde{X}^k.
\end{equation}
The structure constants ${f^k}_{ij}$ and ${{\tilde{f}}^{ij}}_{\; \; \: k}$  are subject to the mixed super Jacobi identities \cite{ER1}
\begin{equation}\label{A5.2}
{f^m}_{jk}{\tilde{f}^{il}}_{\; \; \; \; m}=
{f^i}_{mk}{\tilde{f}^{ml}}_{\; \; \; \; \; j} +
{f^l}_{jm}{\tilde{f}^{im}}_{\; \; \; \; \; k}+ (-1)^{jl}
{f^i}_{jm}{\tilde{f}^{ml}}_{\; \; \; \; \; k}+ (-1)^{ik}
{f^l}_{mk}{\tilde{f}^{im}}_{\; \; \; \; \; j}.
\end{equation}
Now, we assume  $\varepsilon^{+}$ is a $(d_B|d_F)$-dimensional  linear
sub-superspace and $\varepsilon^{-}$ is its orthogonal complement
so that $\varepsilon^{+}+\varepsilon^{-}$ span the whole Lie
superalgebra ${\cal D}$. To determine a dual pair of the
sigma models with the targets $G$ and $\tilde G$, one can consider
the following equation of motion for the mapping $l(\sigma^{+} ,
\sigma^{-})$ from the worldsheet into the Drinfel'd superdouble $D$, \footnote{Note that equation (\ref{A6})
is the field equation coming
from  the first-order
Hamiltonian action for field configurations $l(\tau ,
\sigma)\in D$ ~\cite{klim1}, \cite{klim2}. }
\begin{equation}\label{A6}
< \partial_{\pm}l l^{-1}\;,\;  \varepsilon^{\mp}  >\;=\;0.
\end{equation}
Then, using  equation (\ref{A6}) and the decomposition of
an arbitrary element of $D$ in the vicinity of the unit element of
$D$ as
\begin{equation}\label{A7}
l(\sigma^{+} , \sigma^{-})\; =\; g(\sigma^{+} , \sigma^{-}) \tilde{h}(\sigma^{+}
, \sigma^{-}),\qquad  ~~~~ g\in G, \quad ~~ \tilde{h}\in \tilde{G},
\end{equation}
we obtain
\begin{equation}\label{A8}
<g^{-1} \partial_{\pm}g + \partial_{\pm}\tilde{h}\hspace{0.5mm}
\tilde{h}^{-1} ,g^{-1} \varepsilon^{\mp} g> \;=\; 0,
\end{equation}
for which
\begin{equation}\label{A9}
g^{-1} \varepsilon^{\pm} g\; = \;Span\{X_i \pm E^{\pm}_{ij}(g)
\tilde{X}^j\},
\end{equation}
where  $E^{\pm}:~ {\cal G} \rightarrow {\tilde {\cal G}}$ is a non-degenerate
linear map, and $E^{-}_{ij}= (E^{+}_{ij})^{st} = (-1)^{ij}E^{+}_{ji}$.
Now, inserting relation (\ref{A9}) into (\ref{A8}) and  using  relation (\ref{A10}),
we can write the field equation (\ref{A8}) in the form \cite{ER}
\begin{equation}\label{A11}
(\partial_{+}\tilde{h}\hspace{0.5mm}
\tilde{h}^{-1})_i=(-1)^j\; (g^{-1} \partial_{+}g )^j
E^{+}_{ji}(g),
\end{equation}
\begin{equation}\label{A12}
(\partial_{-}\tilde{h}\hspace{0.5mm}
\tilde{h}^{-1})_i = -E^{+}_{ij}(g) (g^{-1} \partial_{-}g )^j,~~~~~
\end{equation}
where $(\partial_{\pm}\tilde{h}\hspace{0.5mm}
\tilde{h}^{-1})_i$ are right invariant one-forms on $\tilde{G}$ and
satisfy in the following super flat connection equations
\begin{equation}\label{A13}
\partial_{+} (\partial_{-}\tilde{h}\hspace{0.5mm}
\tilde{h}^{-1})_i - \partial_{-} (\partial_{+}\tilde{h}\hspace{0.5mm}
\tilde{h}^{-1})_i -(-1)^{j\hspace{0.5mm}k}\;
{\tilde{f}^{j\hspace{0.5mm}k}}_{\; \; \; \; \; i}\; (\partial_{-}\tilde{h}\hspace{0.5mm}
\tilde{h}^{-1})_j (\partial_{+}\tilde{h}\hspace{0.5mm}
\tilde{h}^{-1})_k = 0.
\end{equation}
However, one can show that equations (\ref{A13}) are the field equations of the sigma model with the action
\begin{equation}\label{A14}
S\;=\;\frac{1}{2}\int\! d\sigma^{+}
d\sigma^{-}\;(-1)^i \;(g^{-1} \partial_{+} g)^i\;{{
E^+_{ij}}}(g) \; (g^{-1} \partial_{-} g)^j,
\end{equation}
in which
\begin{equation}\label{A14.1}
(g^{-1} \partial_{\pm} g)^i := {L_{\pm}}{\hspace{-1mm}^{(l)^i}} = \partial_{\pm} \Phi^{^{\Upsilon}} {_{_\Upsilon }L^{(l)}}^{^{i}},
\end{equation}
are left invariant
one-forms with left derivative. Comparing the  action   (\ref{A14}) with the action (\ref{A1}) we obtain
\begin{equation}\label{A11.1}
{{\cal E}}\hspace{-0.5mm}_{_{\Upsilon \Lambda} }\;=\; (-1)^i\; {L_{_{\Upsilon}}^{(l)}}^i\; {{
E^+_{ij}}}(g)\; {({L^{(l)^{st}}})^j}_{\;\Lambda}.
\end{equation}
Then, using the fact that left invariant
one-forms are dual to the supervector fields, i.e., ${_{_i}V}^{^{\Upsilon}} {_{_\Upsilon }}{L^{(l)}}^{^{j}}={_{_i }}{\delta}^{^{j}}$, one can rewrite  the equations
(\ref{A11}) and (\ref{A12}) in the following form
\begin{equation}\label{A121}
(\partial_{+}\tilde{h}\hspace{0.5mm}
\tilde{h}^{-1})_i\;=\;(-1)^{\Upsilon}\; \partial_{+} \Phi^{^{\Upsilon}}
\;{{\cal E}}\hspace{-0.5mm}_{_{\Upsilon \Lambda} }\;
(V^{st})^{^{\Lambda}}_{~i},~
\end{equation}
\begin{equation}\label{A122}
(\partial_{-}\tilde{h}\hspace{0.5mm}
\tilde{h}^{-1})_i\;=\;-(-1)^{i+\Upsilon}
\;{_{_i}V}^{^{\Upsilon}}\;{{\cal E}}\hspace{-0.5mm}_{_{\Upsilon \Lambda} }\;\partial_{-} \Phi^{^{\Lambda}}.
\end{equation}
Also, one can rewrite the action (\ref{A14}) in the following form
\begin{equation}\label{A16}
S\;=\;\frac{1}{2}\int\!d\sigma^{+}
d\sigma^{-}\;(-1)^i {R_{+}}{\hspace{-1mm}^{(l)^i}}\; {{F}^{^+}}\hspace{-2mm}_{
ij}(g)\;{R_{-}}^{{\hspace{-2mm}(l)^{j}}},
\end{equation}
where ${R_{\pm}}{\hspace{-1mm}^{(l)^i}}$'s are right invariant
one-forms with left derivative and are defined by
\begin{equation}\label{A17}
{R_{\alpha}}{\hspace{-1mm}^{(l)^i}}\;=\;<\partial_{\alpha}g g^{-1}~,~{\tilde X}^i>.
\end{equation}
The matrix ${F^{^+}}\hspace{-2mm}_{
ij}(g)$ of the sigma model is of the form
\begin{equation}\label{A18}
{F^{^+}(g)}\;=\;\Big(\Pi(g)+ ({{E_0}^{\hspace{-1mm}+}})^{-1}(e)\Big)^{-1},~~~~~~~~~~\Pi^{^{ij}}(g)\;=\;(-1)^k b^{ik}(g)\; {(a^{-1})_k}^{~j}(g),
\end{equation}
where ${E_0}^{\hspace{-1mm}+} (e)$ is a constant matrix and $\Pi(g)$ defines the super Poisson structure on the Lie supergroup $G$,
and the sub-matrices $a(g)$ and $b(g)$ are constructed in terms of the bilinear forms as
\begin{equation}\label{A19}
a_i^{\;\;j}(g)\;=~<g^{-1} X_i\; g~,~ {\tilde X}^j>,~~~~~~~~~{b}^{ij}(g)~=~ <g^{-1} \tilde{X}^i g~,~{\tilde X}^j>.
\end{equation}
Equivalently, by using the decomposition
\begin{equation}\label{A20}
l(\sigma^{+} , \sigma^{-})\; =\; {\tilde g}(\sigma^{+} , \sigma^{-}) {h}(\sigma^{+}
, \sigma^{-}),\qquad  ~~~~ \tilde g\in \tilde G, \quad ~~ {h}\in {G},
\end{equation}
and by exchanging $G \leftrightarrow {\tilde G}, {\cal G} \leftrightarrow {\tilde {\cal G}}$
and ${_i{E_{0_{j}}}}^{\hspace{-2mm}+}(e)\leftrightarrow {{\tilde E_0}}^{+^{ij}}(\tilde e)={({E_0}^{\hspace{-1mm}+})^{\hspace{0mm}{-1}}}^{ij}(e)$, the dual sigma model is obtained. Furthermore, we note that  equations (\ref{A7}), (\ref{A11}), (\ref{A12}) and (\ref{A20}) define
the so-called  super Poisson-Lie T-duality
transformations.

Notice that if we take a dual Abelian Lie supergroup (${{\tilde{f}}^{ij}}_{\; \; \: k}=0$) with local coordinates
${\tilde \Phi}_{{k}}$ characterizing the supergroup element $\tilde g$, then, in this case we have
\begin{equation}\label{A18.1}
\Pi^{ij}(g)\;=\;0,~~~~~~~~~~~{\tilde \Pi}^{ij}(\tilde g)\;=\;-{\tilde \Phi}_{{k}}\; {f^k}_{ij},
\end{equation}
recovering,  thus, the super non-Abelian T-duality.
To continue,  we shall present an example of $(2|2)-$dimensional super non-Abelian T-dual sigma models with a curved background
for which the sigma model can be explicitly solved by the super Poisson-Lie T-duality
transformations. This model is obtained from the Drinfel'd superdouble $\left((C^1_1+A)~,~I_{(2|2)}\right)$.

\section{The super non-Abelian T-dual sigma models}

Both the original and the dual geometries of the super Poisson-Lie
T-dualizable sigma models are derived from the so-called Drinfel'd superdouble which is a Lie supergroup.
We shall construct a dualizable sigma model on the $(2|2)$-dimensional Lie supergroup $(C^1_1+A)$ when the dual Lie supergroup
is the $(2|2)$-dimensional Abelian Lie supergroup, hence, the super Poisson-Lie T-duality reduces to the
super non-Abelian T-duality. To this end, we first introduce the Lie superalgebra $({\cal C}^1_1+{\cal A})$ \cite{B} of the Lie supergroup $(C^1_1+A)$.
As mentioned in introduction Section, the Lie superalgebra $({\cal C}^1_1+{\cal A})$ possesses two bosonic generators $X_1$ and $X_2$ along with two
fermionic ones $X_3$ and $X_4$. These four generators obey
the following set of non-trivial (anti)commutation relations \cite{B}
\begin{equation}\label{B1}
[X_1 , X_2]=X_2,~~~~~[X_1 , X_3]=X_3,~~~~~\{X_3 , X_4\}=X_2.
\end{equation}
The Lie superalgebra of the
Drinfel'd superdouble which we refer to as  $\left(({\cal C}^1_1+{\cal A})\;,\;{\cal I}_{(2|2)}\right)$ is denoted by generators $\{ X_1, X_2, {\tilde
X}^1, {\tilde X}^2; X_3, X_4, {\tilde X}^3, {\tilde X}^4 \}$ and the
following nonzero (anti)-\\commutation relations
$$
[X_1 , X_2]=X_2,~~~~~[X_1 , X_3]=X_3,~~~~~\{X_3 , X_4\}=X_2,~~~~~~~
$$
\vspace{-4mm}
$$
[X_1 , {\tilde X}^2]=-{\tilde X}^2,~~~~~[X_1 , {\tilde X}^3]=-{\tilde X}^3,~~~~[X_2 , {\tilde X}^2]={\tilde X}^1,~~~
$$
\vspace{-4mm}
\begin{equation}\label{B2}
[X_3 , {\tilde X}^2]=-{\tilde X}^4,~~~~[X_4 , {\tilde X}^2]=-{\tilde X}^3,~~~~\{ X_3 , {\tilde X}^3\}=-{\tilde X}^1,~
\end{equation}
where $\{ {\tilde X}^1, {\tilde X}^2\}$ and $\{{\tilde X}^3 , {\tilde X}^4\}$ are the respective bosonic  and fermionic generators of the Abelian Lie superalgebra.

\subsection{The original model}\label{s3.1}
The super Poisson-Lie T-dual sigma models are usually expressed in terms of supergroup coordinates. These coordinates are given
by the Lie supergroup structure and follow from the possibility to express the elements of the Lie supergroup
as a product of elements of one-parametric sub-(super)groups.
In order to construct the  corresponding original sigma model with
the Lie supergroup $({ C}^1_1+{ A})$ as the target space, we use the supergroup coordinates  $\Phi^{^{\Upsilon}}=\{x, y; \psi, \chi\}$, for which
the elements of the Lie supergroup $G$ are parametrized  as
\begin{equation}\label{B3}
g\;=\; e^{\chi X_4} ~e^{y X_2}~ e^{x X_1} ~  e^{\psi X_3},
\end{equation}
where we have, here, assumed that $g$ is a mapping from $\Sigma$ into $G$. The fields  $x(\tau , \sigma)$ and $y(\tau , \sigma)$ are bosonic while $\psi(\tau , \sigma)$ and $\chi(\tau , \sigma)$ are fermoinic fields.
Inserting the above parametrization in
(\ref{A17}) and  (\ref{A14.1}),  ${{R}^{(l)}_\pm}^i$'s  and  ${{L}^{(l)}_\pm}^i$'s are, respectively,  found to be
\begin{eqnarray}
{{R}^{(l)}_\pm}^{^{X_1}} & = & {\partial_\pm}x,~~~~~~~~~~~~~~
{{R}^{(l)}_\pm}^{^{X_2}} = -{\partial_\pm}x ~y+{\partial_\pm}y+{\partial_\pm}\psi e^x \chi,\nonumber\\
{{R}^{(l)}_\pm}^{^{X_3}} & = & -{\partial_\pm}\psi e^x,~~~~~~~~
{{R}^{(l)}_\pm}^{^{X_4}} = -{\partial_\pm}\chi, \label{B4}
\end{eqnarray}
and
\begin{eqnarray}
{{L}^{(l)}_\pm}^{^{X_1}} & = & {\partial_\pm}x,~~~~~~~~~~~~~~~~~~~~~
{{L}^{(l)}_\pm}^{^{X_2}} = {\partial_\pm}y ~e^{-x}- {\partial_\pm}\chi ~\psi,\nonumber\\
{{L}^{(l)}_\pm}^{^{X_3}} & = & -{\partial_\pm} x~ \psi -{\partial_\pm}\psi ,~~~~~~
{{L}^{(l)}_\pm}^{^{X_4}} = -{\partial_\pm}\chi. \label{B4.1}
\end{eqnarray}
In this case, $\Pi (g)=0$ because the dual Lie supergroup is Abelin. Hence, choosing the sigma model matrix $E_0^+(e)$ at the unit element
of the Lie supergroup $({ C}^1_1+{ A})$ as
\begin{equation}\label{B5}
E_{0_{ij}}^+(e)=\left( \begin{tabular}{cccc}
              $0$ & $K$ & $0$ & $0$\\
              $K$ & $0$ & $0$ & $0$ \\
              $0$ & $0$ & $0$ & $K$ \\
              $0$ & $0$ & $-K$ & $0$\\
                \end{tabular} \right),
\end{equation}
where $K$ is a non-zero constant, and  using the first equation of (\ref{A18}), the original sigma model action is worked out as follows:
\begin{eqnarray}\label{B6}
S  =  \frac{K}{2} \int d\sigma^{+} d\sigma^{-}\;\Big (\hspace{-2mm}&-& \hspace{-2mm}2 y\partial_{+} x \partial_{-} x+
\partial_{+} x
\partial_{-} y + \partial_{+} y
\partial_{-} x \nonumber\\
& -& \hspace{-1mm}  \partial_{+} x e^x \chi \partial_{-} \psi  +\partial_{+} \psi e^x \chi  \partial_{-} x
-\partial_{+} \psi e^x \partial_{-} {\chi} +\partial_{+} \chi e^x \partial_{-} {\psi} \Big).
\end{eqnarray}
This model is not only
Ricci flat and torsionless but also is flat in the sense that its Riemann tensor vanishes. Thus, we shall
 deal with a model having  ${\cal E}_{_{\Upsilon\Lambda}}=G_{_{\Upsilon\Lambda}}$ so that
\begin{equation}\label{B7}
G_{_{\Upsilon\Lambda}}(\Phi)=
\left(\begin{array}{cccc}
-2K y & K & -Ke^x \chi&0\\
K & 0 & 0&0\\
-Ke^x \chi & 0 & 0 & Ke^x \\
0 & 0 & -Ke^x &0
\end{array}\right).
\end{equation}

\subsection{The dual model}\label{s3.2}

In the same way to construct the dual sigma model, we parametrize
the corresponding Lie supergroup (Abelian Lie supergroup) with
bosonic coordinates $\{\tilde x, \tilde y\}$ and fermionic ones
$\{\tilde \psi, \tilde \chi\}$ so that its elements can be
written as:
\begin{equation} \label{B8}
\tilde g\;=\; e^{\tilde \chi {\tilde X}^4}  e^{\tilde y {\tilde
X}^2} e^{\tilde x {\tilde X}^1}  e^{\tilde \psi{\tilde X}^3}.
\end{equation}
Since the dual Lie supergroup is Abelian, so the components of
the right invariant one-forms are simply calculated to be
${{{_{_{\Upsilon}}{\tilde R}}_{\pm i}}^{\hspace{-2mm}(l)}}=
{{_{_{\Upsilon}}{\tilde \delta}}_{i}}$ (${{_{_{\Upsilon}}{\tilde \delta}}_{i}}$
is the ordinary delta function). Using the structure constants of the Lie superalgebra
$({\cal C}^1_1+{\cal A})$ (relation (\ref{B1})) and then  utilizing  the second equation of (\ref{A18.1}),
the super Poisson structure on the dual Lie supergroup is found to be
\begin{equation}\label{B9}
{\tilde \Pi}_{ij}(\tilde g) \;=\;\left( \begin{tabular}{cccc}
              $0$ & $-\tilde y$ & $-{\tilde \psi}$ & $0$\\
              $\tilde y$ & $0$ & $0$ & $0$ \\
              ${\tilde \psi}$ & $0$ & $0$ & $-\tilde y$ \\
              $0$ & $0$ & $-{\tilde y}$ & $0$\\
                \end{tabular} \right).
\end{equation}
It is quite interesting that the super Poisson structure is
superinvertible, because one can use the relation
\begin{equation}\label{B10}
{\tilde \omega}^{^{\Upsilon\Lambda}}(\tilde g) = (-1)^i ({\tilde
L}^{(l)^{-st}})^{^{\Upsilon i}} {\tilde \Pi}_{ij}(\tilde g)
({\tilde L}^{(l)^{-1}})^{^{j \Lambda}},
\end{equation}
to construct a even supersymplectic two-form\footnote{Note that in terms of the coordinate basis $\{d {\tilde \Phi}^{^{\Lambda}} \}$, a two-form ${\tilde \omega}$ is defined as ${\tilde \omega} =
\frac{(-1)^{^{\Upsilon \Lambda }}}{2} ~{\tilde \omega}_{_{\Upsilon \Lambda}} d {\tilde \Phi}^{^{\Upsilon}} \wedge d {\tilde \Phi}^{^{\Lambda}} $ (for more detailed description see D{\tiny E}WITT's book \cite{D}). } ${\tilde
\omega}$ on the dual supergroup supermanifold
${\tilde G}=I_{(2|2)}$ as follows:
\begin{equation}\label{B10.1}
{\tilde \omega}=\frac{1}{\tilde y}\Big( d{\tilde x} \wedge {d\tilde y} + \frac{{\tilde \psi}} {{\tilde y} }d{\tilde y} \wedge {d\tilde \chi}-
d{\tilde \psi} \wedge {d\tilde \chi} \Big).
\end{equation}
It is easy to show that ${d\tilde \omega} =0$. Moreover, one can write the ${\tilde \omega}$ as the exterior derivative of one-form
\begin{equation}\label{B10.111}
{\tilde \eta} = f({\tilde x}) d{\tilde x}  + \frac{\tilde x}{\tilde y} {d\tilde y} + {\cal{C}}{\tilde \psi} d{\tilde \psi} -
\frac{\tilde \psi}{\tilde y} {d\tilde \chi},
\end{equation}
where $f({\tilde x})$ is an arbitrary function of $\tilde x$ and ${\cal{C}}$ is an even, real constant.
Let us, now, write the action of dual sigma model. Substituting (\ref{B9}) in the first
equation of (\ref{A18}) (equation  (\ref{A18}) in the dual form)  and then using the fact that
$({\tilde E}_0^{\hspace{1mm}+})^{-1}(\tilde e) =
E_0^{\hspace{1mm}+}(e)$, the dual sigma model action is obtained
to be
\begin{eqnarray}\label{B11}
\tilde S  =  \frac{1}{2} \int d\sigma^{+} d\sigma^{-}\;
(\frac{1}{K^2-{\tilde y}^2}) \Big [\hspace{-2mm}&& \hspace{-2mm}
(K-\tilde y)(\partial_{+} {\tilde x}
\partial_{-} {\tilde y}-\partial_{+} {\tilde \psi}
\partial_{-} {\tilde \chi})+(K+\tilde y)(\partial_{+} {\tilde y}
\partial_{-} {\tilde x}+\partial_{+} {\tilde \chi}
\partial_{-} {\tilde \psi}) \nonumber\\
~& -&\hspace{-1mm}\partial_{+} {\tilde y}~ {\tilde \psi}~
\partial_{-} {\tilde \chi} -\partial_{+} {\tilde \chi}~{\tilde \psi}
\partial_{-} {\tilde y} \Big].
\end{eqnarray}
Comparing  the above action with the sigma model action
(\ref{A1}), we can elicit the  metric ${\tilde G}_{\Upsilon
\Lambda}$ and the field ${\tilde B}_{\Upsilon \Lambda}$ as
{\small \begin{equation}\label{B12}
{\tilde G}_{\Upsilon \Lambda} \;=\; \frac{K}{K^2-{\tilde
y}^2}\left( \begin{tabular}{cccc}
              $0$ & $1$ & $0$ & $0$\\
              $1$ & $0$ & $0$ & $0$ \\
              $0$ & $0$ & $0$ & $1$ \\
              $0$ & $0$ & $-1$ & $0$\\
                \end{tabular} \right),~~{\tilde B}_{\Upsilon \Lambda} \;=\;\frac{1}{K^2-{\tilde y}^2}\left( \begin{tabular}{cccc}
              $0$ & $-\tilde y$ & $0$ & $0$\\
              $\tilde y$ & $0$ & $0$ & $-{\tilde \psi}$ \\
              $0$ & $0$ & $0$ & $-\tilde y$ \\
              $0$ & ${\tilde \psi}$ & $-\tilde y$ & $0$\\
                \end{tabular} \right).
\end{equation}}
The investigated  dual model has non-zero anti-supersymmetric
part ${\tilde B}_{_{\Upsilon\Lambda}}$ of the tensor ${\tilde
{\cal E}}_{_{\Upsilon\Lambda}}$, but it is flat in the sense that its scalar
curvature vanishes.

\section{Flat supercoordinates}

Riemannian  coordinates  on supermanifolds are one type of  coordinates where the sigma models live.
In these coordinates, the metric on supermanifold have a special simple form. The Riemannian coordinates of the flat metrics
are, here,  called \emph{flat supercoordinates}.

We know that the components of the connection do not transform as the components of a mixed tensor field. Thus,
when the transform is from one coordinate basis to another, then,  ${\Gamma}^{^{\Upsilon}}_{_{~\Lambda\Delta}}$ transforms as follows \cite{D}:
\begin{equation}\label{c1}
{\Gamma}^{^{\Upsilon}}_{_{\;~\Lambda \Delta}} \;=\; (-1)^{^{C(B+\Lambda)}}\;  \Big(\frac{ \Phi ^{^{\Upsilon}}\overleftarrow{\partial}}{\partial
\Omega^{^{A}}} \Big){\bar{\Gamma}}^{^{A}}_{_{\;B C}} \; \Big(\frac{\Omega^{^{B}} \overleftarrow{\partial}}{\partial
\Phi ^{^{\Lambda}} } \Big)\Big(\frac{\Omega^{^{C}} \overleftarrow{\partial}}{\partial
\Phi ^{^{\Delta}} } \Big)+\Big(\frac{ \Phi ^{^{\Upsilon}}\overleftarrow{\partial}}{\partial
\Omega^{^{A}}} \Big)\Big((\frac{\Omega^{^{A}} \overleftarrow{\partial}}{\partial
\Phi ^{^{\Lambda}}})\frac{ \overleftarrow{\partial}}{\partial
\Phi ^{^{\Delta}}} \Big).
\end{equation}
For finding the flat supercoordinates $\Omega^{^{A}}$ (here, $\Omega^{^{A}}=\{\Omega^{^{1}}, \Omega^{^{2}}; \Omega^{^{3}}, \Omega^{^{4}}\}$, where $\Omega^{^{1}}$
and  $\Omega^{^{2}}$ are bosonic coordinates while $\Omega^{^{3}}$
and  $\Omega^{^{4}}$ are fermionic ones), we shall use formula (\ref{c1}). In these supercoordinates the metric becomes constant and, therefore,
the elements of ${\bar{\Gamma}}^A_{\;B C} $ vanish.  Then, we get the following system of partial differential equations (PDEs) for $\Omega(\Phi)$
\begin{equation}\label{c2}
(-1)^{^{\Upsilon + \Upsilon B}}\;\frac{ \overrightarrow{\partial} \Omega^{^{B}}}{\partial
\Phi ^{^{\Upsilon}} } {\Gamma}^{^{\Upsilon}}_{_{\;~\Lambda \Delta}} \;=\;(-1)^{^{(\Lambda + \Delta)(1+ B)}}\;\frac{ \overrightarrow{\partial}}{\partial
\Phi ^{^{\Lambda}} }\Big(\frac{ \overrightarrow{\partial} \Omega^{^{B}}}{\partial
\Phi ^{^{\Delta}} } \Big).
\end{equation}
The possibility  to solve the above equation depends on the form of $ {\Gamma}^\Upsilon_{\;~\Lambda \Delta}$.
For the metric given by the relation (\ref{B7}) we find the general explicit solution with the initial condition that will produce
the Riemannian  coordinates.  The initial condition for producing the  flat supercoordinates, in which the metric
requires the constant form $\bar{G}(\Omega)=G(\Phi=0)$, is given by
\begin{equation}\label{c3}
{\frac{\Omega^{^{A}} \overleftarrow{\partial}}{\partial
\Phi ^{^{\Upsilon}} }|}_{\Phi=0}\;=\;{\delta^{^{A}}}_{_{\Upsilon}}.
\end{equation}
To continue, we shall present the solution of equations
(\ref{c2}) in detail for the metric (\ref{B7}).
\section{Solving the equations for flat supercoordinates}

In this section we find the transformation of supercoordinates that makes the metric constant and use it for solving
the models. As mentioned in subsection (\ref{s3.1}),  the model given by the metric (\ref{B7}) lives in the flat background in the sense that its Riemann
tensor vanishes. In spite of the fact that the metric is flat, the $ {\Gamma}^\Upsilon_{\;~\Lambda \Delta}$ are not zero, so, it is not easy
to find  the  coordinates $\Phi^{^{\Upsilon}}(\tau , \sigma)$ that solve the equations of motion given by the action (\ref{A1}).
After the use of the relation (\ref{A2}) to calculate the components of the connection  for the metric (\ref{B7}), equations (\ref{c2}) then read
\begin{eqnarray}
-\frac{\partial \Omega^{^{A}}}{\partial y}&=&\frac{\partial^{2}\Omega^{^{A}}}{\partial x\partial y},\label{d1}\\
-e^x\;\frac{\partial \Omega^{^{A}}}{\partial y}&=&\frac{\partial^{2}\Omega^{^{A}}}{\partial \psi \partial \chi},\label{d2}\\
\frac{\partial \Omega^{^{A}}}{\partial x } +2y \; \frac{\partial \Omega^{^{A}}}{\partial y}&=&\frac{\partial^{2}\Omega^{^{A}}}{\partial x^2},\label{d3}\\
\frac{\partial \Omega^{^{A}}}{\partial y }\; e^x \chi - (-1)^A \; \frac{\partial \Omega^{^{A}}}{\partial \psi}&=&
-(-1)^A\;\frac{\partial^{2}\Omega^{^{A}}}{\partial x  \partial \psi },\label{d4}\\
\frac{\partial^{2}\Omega^{^{A}}}{\partial y^2 }=0,~~ \frac{\partial^{2}\Omega^{^{A}}}{\partial \psi^2 }=0,~~
 \frac{\partial^{2}\Omega^{^{A}}}{\partial \chi^2 }&=&0,\label{d5}\\
\frac{\partial^{2}\Omega^{^{A}}}{\partial x  \partial \chi }=0,~ ~\frac{\partial^{2}\Omega^{^{A}}}{\partial y  \partial \psi }=0,~~
 \frac{\partial^{2}\Omega^{^{A}}}{\partial y  \partial \chi }&=&0,\label{d6}
\end{eqnarray}
where $(-1)^{^{A}}=\left\{\begin{array}{ll}1&{\rm } \;A =1, 2\\
-1&{\rm } \;A =3, 4\\\end{array} \right.$. The above equations can be solved and the general solution is
\begin{eqnarray}
\Omega^{^{1}}&=& a_{_{0}}  + a_{_{1}} (ye^{-x} + \psi \chi) + a_{_{2}} e^x,\nonumber\\
\Omega^{^{2}}&=& b_{_{0}}  + b_{_{1}}(ye^{-x} + \psi \chi) + b_{_{2}} e^x,\nonumber\\
\Omega^{^{3}}&=& {\alpha}_{_{0}}  + c_{_{1}} \chi  + c_{_{2}} e^x \psi,\nonumber\\
\Omega^{^{4}}&=& {\beta}_{_{0}}  + d_{_{1}} \chi  + d_{_{2}} e^x \psi,\label{d7}
\end{eqnarray}
where the integration constants ${\alpha}_{_{0}}$ and ${\beta}_{_{0}}$ are real  $a-$numbers (odd real constants)
while the rest of constants are real $c-$numbers (even real constants) \cite{D}.
Imposing the initial condition (\ref{c3}), the integration constants will be determined. Then, the solution (\ref{d7}) takes the following form
\begin{eqnarray}
\Omega^{^{1}}&=& a_{_{0}}  +  e^x,\nonumber\\
\Omega^{^{2}}&=& b_{_{0}}  + ye^{-x} + \psi \chi,\nonumber\\
\Omega^{^{3}}&=& {\alpha}_{_{0}}  + e^x \psi,\nonumber\\
\Omega^{^{4}}&=& {\beta}_{_{0}}  + \chi.\label{d8}
\end{eqnarray}
In the following, for simplicity we will assume that $ {\alpha}_{_{0}} = {\beta}_{_{0}} =0$.
The transformation (\ref{d8}) transforms the metric (\ref{B7}) into constant one
\begin{equation}\label{d9}
\bar{G}_{_{AB}}(\Omega)=
\left(\begin{array}{cccc}
0 & K & 0 & 0\\
K & 0 & 0 & 0\\
0 & 0 & 0 & K \\
0 & 0 & -K&0
\end{array}\right).
\end{equation}
Thus, the action corresponding to the metric (\ref{d9}) is given by
\begin{eqnarray}
\bar{S}  &=&  \frac{1}{2} \int d\sigma^{+} d\sigma^{-}\;(-1)^A\;
\partial_{+} \Omega^A\; \bar{G}_{_{AB}}\;\partial_{-} \Omega^B \nonumber\\
&=& \frac{K}{2} \int d\sigma^{+} d\sigma^{-}\;\Big(\partial_{+} \Omega^1 \partial_{-} \Omega^2
+\partial_{+} \Omega^2 \partial_{-} \Omega^1 - \partial_{+} \Omega^3 \partial_{-} \Omega^4 + \partial_{+} \Omega^4 \partial_{-} \Omega^3 \Big).\label{d10}
\end{eqnarray}
Now, we show that the transformation (\ref{d8}) is indeed a classical supercanonical transformation.

\subsection{Supercoordinates transformation as a  supercanonical transformation}
This subsection begins with the calculation of the  momentums  corresponding to the actions (\ref{B6}) and (\ref{d10}) and  the transformation between them.
To this end, we first write  Lagrangians  of the  actions (\ref{B6}) and (\ref{d10}) in the worldsheet coordinates.
Using the definition of the conjugate momentum of $\Phi^{^{\Upsilon}}$ on the supermanifolds as
\begin{eqnarray}\label{d11}
P_{_{\Phi^{^{\Upsilon}}}}\;=\; \frac{L \overleftarrow{\delta}}{ \delta (\partial_{\tau}
\Phi ^{^{\Upsilon}} )},
\end{eqnarray}
the corresponding moments are then found to be
\begin{eqnarray}
P_{_{x}}& =& K(-2y~\partial_{\tau} x + \partial_{\tau} y-e^x \chi\; \partial_{\tau} \psi),~~~~\;~ P_{_{y}}~=~K \partial_{\tau} x,     \nonumber\\
P_{_{\psi}}&=& K(-e^x\;\partial_{\tau} x~\chi + e^x  \partial_{\tau} \chi), ~~~~~~~~~~~~~~P_{_{\chi}}~=~-K e^x \;\partial_{\tau} \psi,    \label{d12}
\end{eqnarray}
for the model (\ref{B6}) and
\begin{eqnarray}
\bar{P}_{_{\Omega^{^{1}}}}& =& K \partial_{\tau} {\Omega^{^{2}}},~~~~~~~~~~
\bar{P}_{_{\Omega^{^{2}}}}~=~K \partial_{\tau} {\Omega^{^{1}}},     \nonumber\\
\bar{P}_{_{\Omega^{^{3}}}}&=& K \partial_{\tau} {\Omega^{^{4}}}, ~~~~~~~~~~
\bar{P}_{_{\Omega^{^{4}}}}~=~-K \partial_{\tau} {\Omega^{^{3}}},   \label{d13}
\end{eqnarray}
for the model (\ref{d10}). The transformation (\ref{d8}) is a transformation between supercoordinates $\Phi^{^{\Upsilon}}$ and $\Omega^{^{A}}$.
Utilizing the transformation (\ref{d8}) and contracting relations
(\ref{d12}) and  (\ref{d13}), the  transformation between momentums is  also obtained to be of the form
\begin{eqnarray}
\bar{P}_{_{\Omega^{^{1}}}} & =& e^{-x} (P_{_{x}} + y P_{_{y}}+ \psi\;P_{_{\psi}}) + \psi \chi P_{_{y}},~~~~~~~
\bar{P}_{_{\Omega^{^{2}}}} ~=~ e^x ~ P_{_{y}},    \nonumber\\
\bar{P}_{_{\Omega^{^{3}}}} & =& e^{-x}\; P_{_{\psi}} +  \chi P_{_{y}},~~~~~~~~~~~~~~~~~~~~~~~~~~~~~~
\bar{P}_{_{\Omega^{^{4}}}} ~=~ -e^x \psi\; P_{_{y}}+P_{_{\chi}}.   \label{d14}
\end{eqnarray}
It can be useful to comment on the fact that the graded Poisson brackets are preserved when equations (\ref{d14}) are combined with (\ref{d8}).
The basic equal-time graded Poisson brackets for the pair of supercanonical variables $(\Phi^{^{\Upsilon}} , P_{_{\Lambda}})$ are given by
\begin{eqnarray}
\{\Phi^{^{\Upsilon}} ~,~ P_{_{\Lambda}} \} & =& {{\delta}^{^{\Upsilon}}}_{_{\Lambda}}~ \delta(\sigma-\sigma'),\nonumber\\
\{\Phi^{^{\Upsilon}} ~,~ \Phi^{^{\Lambda}} \} & =& \{P_{_{\Upsilon}} ~,~ P_{_{\Lambda}} \}~=~0.   \label{d14.1.1}
\end{eqnarray}
It is understood that the first one on the bracket is always evaluated at $\sigma$  and the second one at $\sigma'$.
Using equations (\ref{d14}) and (\ref{d8}) together with (\ref{d14.1.1}) one can show that  the  equal-time graded Poisson brackets for the pair of supercanonical variables $(\Omega^{^{A}} , {\bar{P}}_{_{B}})$ are also preserved; that is,
\begin{eqnarray}
\{\Omega^{^{A}} ~,~ {\bar{P}}_{_{B}} \} & =& {{\delta}^{^{A}}}_{_{B}}~ \delta(\sigma-\sigma'),\nonumber\\
\{\Omega^{^{A}} ~,~ \Omega^{^{B}} \} & =& \{{\bar{P}}_{_{A}} ~,~ {\bar{P}}_{_{B}}\}~=~0.   \label{d14.1.2}
\end{eqnarray}
Let us now  to obtain the Hamiltonian corresponding to the action (\ref{d10}). Utilizing relation (\ref{d13}) we get
\begin{eqnarray}
\bar{{\cal H}}(\Omega)~=~ K \Big(\partial_{\tau} \Omega^{^{1}}\;\partial_{\tau} \Omega^{^{2}}+
\partial_{\sigma} \Omega^{^{1}}\;\partial_{\sigma} \Omega^{^{2}}-
\partial_{\tau} \Omega^{^{3}}\;\partial_{\tau} \Omega^{^{4}}-
\partial_{\sigma} \Omega^{^{3}}\;\partial_{\sigma} \Omega^{^{4}}\Big).  \label{d15}
\end{eqnarray}
Under the transformation  (\ref{d8}), the  Hamiltonian (\ref{d15}) turns into
\begin{eqnarray}
{{\cal H}(\Phi)}~=~ K \Big( &&\hspace{-4mm} \partial_{\tau} x\;\partial_{\tau} y+
\partial_{\sigma} x\;\partial_{\sigma} y-
y \partial_{\tau} x\;\partial_{\tau} x- y \partial_{\sigma} x\;\partial_{\sigma} x  \nonumber\\
~~&-&\hspace{-1mm}  e^x\;
\partial_{\tau} x \;\chi\; \partial_{\tau} \psi - e^x\;
\partial_{\sigma} x \;\chi\; \partial_{\sigma} \psi-
e^x\;\partial_{\tau} \psi\;\partial_{\tau} \chi-e^x\;\partial_{\sigma} \psi\;\partial_{\sigma} \chi \Big).  \label{d16}
\end{eqnarray}
One can simply show that the Hamiltonian (\ref{d16}) is nothing but the Hamiltonian corresponding to the action
(\ref{B6}). Thus, we proved that the Hamiltonians corresponding to the actions
(\ref{B6}) and (\ref{d10}) are equal; that is, under the transformation of supercoordinates (\ref{d8}), one goes from
$\bar{{\cal H}}(\Omega)$ to ${{\cal H}(\Phi)}$ and vice versa, hence, proving that the  transformation  (\ref{d8}) or  (\ref{d14})  is indeed  a supercanonical transformation.

A bit surprisingly, one can show that under the linear  transformation
\begin{eqnarray}
\Omega'^{^{1}}&=& \frac{1}{\sqrt{2}}\; (K\Omega^{^{2}}-\Omega^{^{1}}),\nonumber\\
\Omega'^{^{2}}&=& \frac{1}{\sqrt{2}}\; (K\Omega^{^{2}}+\Omega^{^{1}}),\nonumber\\
\Omega'^{^{3}}&=& -\Omega^{^{3}},\nonumber\\
\Omega'^{^{4}}&=& -K\Omega^{^{4}},\label{d17}
\end{eqnarray}
the action (\ref{d10}) reduces to the following action
\begin{eqnarray}\label{d18}
{S'}  &=&  \frac{1}{2} \int d\sigma^{+} d\sigma^{-}\;\Big(\partial_{+} \Omega'^{\mu}\; \partial_{-} \Omega'^{\nu} \eta_{\mu\nu}
+\partial_{+} \Omega'^{\alpha}\; \partial_{-} \Omega'^{\beta} ~G_{\beta \alpha} \Big),
\end{eqnarray}
where the indices $\mu, \nu$ and $\alpha, \beta$ range over the values $1, 2$ and $3, 4$, respectively. So, $\eta_{\mu\nu}=diag(-1~,~1)$ denotes
the Minkowski metric of the bosonic directions, and the metric of  the fermionic directions is
\begin{equation}\label{d19}
G_{\beta \alpha}\;=\;
\left(\begin{array}{cc}
0 & 1 \\
-1 & 0
\end{array}\right).
\end{equation}
The action (\ref{d18}) is nothing but the Polyakov action on a $(2|2)$-dimensional flat supermanifold \cite{Tokunaga}. Meanwhile, one can show that
the linear  transformation  (\ref{d17}) is also a supercanonical transformation.

\section{Calculation  of the dual solution using the super Poisson-Lie T-duality  transformations}
The relation between the solutions $\Phi^{^{\Upsilon}}(\sigma^+, \sigma^-)$
of the original model (\ref{B6}) and ${\tilde \Phi}^{^{\Upsilon}}(\sigma^+, \sigma^-)$
of the dual model (\ref{B11}) is given by two possible decompositions (\ref{A7}) and (\ref{A20}), in which the $\tilde h$ satisfies the equations
(\ref{A121}) and (\ref{A122}). Therefore, for a solution ${\Phi}^{^{\Upsilon}}(\sigma^+, \sigma^-)$
of the sigma model we must find ${\tilde h}(\sigma^+, \sigma^-)$, i.e., solve the system of PDEs (\ref{A121}) and (\ref{A122}).
Then, using the transformations  (\ref{A7}) and (\ref{A20}) we find the solutions ${\tilde \Phi}^{^{\Upsilon}}(\sigma^+, \sigma^-)$
of the dual sigma model. To this end, by introducing
\begin{eqnarray}
\Omega^{^{a}}&=& W^a(\sigma^+) + U^a(\sigma^-),~~~~~~~~a=1, 2,\label{d20}\\
\Omega^{^{\alpha}}&=& \eta^{\alpha}(\sigma^+) + \xi^{\alpha}(\sigma^-),~~~~~~~~~~\alpha=3, 4,\label{d21}
\end{eqnarray}
we transform the equations of motion to the wave equations. In relations (\ref{d20}) and (\ref{d21}), the
$W^a(\sigma^+)$ and the $U^a(\sigma^-)$ are even, arbitrary functions  while the  $\eta^{\alpha}(\sigma^+)$ and
the  $\xi^{\alpha}(\sigma^-)$ are odd, arbitrary ones. Thus, functions  $\Phi^{^{\Upsilon}}(\sigma^+, \sigma^-)$
follow from  (\ref{d8}), (\ref{d20}) and (\ref{d21})
\begin{eqnarray}
e^x&=& W^1(\sigma^+) + U^1(\sigma^-)-a_{_{0}} , \nonumber\\
y&=& \Big(W^1(\sigma^+) + U^1(\sigma^-)- a_{_{0}}   \Big) \Big(W^2(\sigma^+) + U^2(\sigma^-)- b_{_{0}}  \Big)\nonumber\\
&&~~~~~~~~~~~~~~~~~~~~~~~~~-\Big(\eta^{3}(\sigma^+) + \xi^{3}(\sigma^-)\Big)
\Big(\eta^{4}(\sigma^+) + \xi^{4}(\sigma^-)\Big),\nonumber\\
\psi&=&\frac{\eta^{3}(\sigma^+) + \xi^{3}(\sigma^-)}{ W^1(\sigma^+) + U^1(\sigma^-)-a_{_{0}} },\nonumber\\
\chi&=&\eta^{4}(\sigma^+) + \xi^{4}(\sigma^-).\label{d22}
\end{eqnarray}
Now one can show that the above functions satisfy the following equations
\begin{eqnarray}
0&=&\partial _+ \partial _-  x +  \partial _- {x} ~\partial _+ x,\nonumber\\
0&=&\partial _+ \partial _- y + 2 y \partial _- {x} ~\partial _+ x -\partial _- {y} ~\partial _+ x  - \partial _- {x} ~\partial _+ y \nonumber\\
&&+e^x \chi(\partial _- {\psi} ~\partial _+ x + \partial _- {x} ~\partial _+ \psi)+ e^x(\partial _- {\psi} ~\partial _+ \chi - \partial _- {\chi} ~\partial _+ \psi),\nonumber\\
0&=&\partial _+ \partial _- \psi + \partial _- {x}~\partial _+ {\psi}+\partial _- {\psi} \partial _+ x,\nonumber\\
0&=&\partial _+ \partial _-  \chi.\label{d22.1}
\end{eqnarray}
Note that equations (\ref{d22.1}) are the equations of motion of action (\ref{B6}) which have been obtained by using relation (\ref{A4}).
Thus, we could find the solution of the equations of motion of the original model by the use of the flat supercoordinates transformation.
In the following to find ${\tilde h}(\sigma^+, \sigma^-)$, we solve the system of PDEs (\ref{A121}) and (\ref{A122}). To make up the right hand sides
of the equations  (\ref{A121}) and (\ref{A122})  one must use (\ref{B4.1}) to calculate ${_{_i}V}^{^{\Upsilon}}$'s. Then, using (\ref{B7})
the right hand sides  read
\begin{eqnarray}
(\partial_{\pm}\tilde{h}\hspace{0.5mm}
\tilde{h}^{-1})_1&=& \pm K\Big[-2(y\partial_{\pm}x + \psi \chi \partial_{\pm} e^x) +  \partial_{\pm}y +\partial_{\pm}  (e^x \psi \chi) \Big], \nonumber\\
(\partial_{\pm}\tilde{h}\hspace{0.5mm}\tilde{h}^{-1})_2&=& \pm K \partial_{\pm}e^x, \nonumber\\
(\partial_{\pm}\tilde{h}\hspace{0.5mm}\tilde{h}^{-1})_3&=& \pm K (\chi \partial_{\pm}e^x -e^x \partial_{\pm} \chi),\nonumber\\
(\partial_{\pm}\tilde{h}\hspace{0.5mm}\tilde{h}^{-1})_4&=& \pm K \partial_{\pm}(e^x\; \psi).\label{d23}
\end{eqnarray}
On the other hand, if we parametrize  elements of the Abelian Lie supergroup $\tilde G$ with bosonic coordinates $\{{\tilde h}_1, \;{\tilde h}_2\} $ and fermionic
ones $\{{\tilde h}_3, \;{\tilde h}_4\}$ as
\begin{eqnarray}
\tilde h\;=\; e^{{\tilde h}_4 {\tilde X}^4} \; e^{{\tilde h}_2 {\tilde X}^2}\; e^{{\tilde h}_1 {\tilde X}^1}   \; e^{{\tilde h}_3 {\tilde X}^3}, \label{d24}
\end{eqnarray}
then, the left hand sides are just $\partial_{\pm}\tilde{h}_i$.  Substituting the solution
$\Phi^{^{\Upsilon}}(\sigma^+, \sigma^-)$ given by  (\ref{d22}) into (\ref{d23}), the system of PDEs (\ref{A121}) and (\ref{A122})
can be solved and the general solution is\footnote{Here, $Q_{\pm}$ and $W_+^a$ have been used instead of $Q(\sigma^+, \sigma^-)$ and $W^a(\sigma^+)$, respectively. The other notations follow this rule, too. }
\begin{eqnarray}
{\tilde h}_1&=&K\Big[Q_{\pm}+   U_-^1 W_+^2-  U_-^2 W_+^1+b_{_{0}} \Big(W_+^1 - U_-^1\Big)-a_{_{0}} \Big(W_+^2 - U_-^2\Big)\Big]+ c_{_{0}} , \nonumber\\
{\tilde h}_2&=& K\Big(W_+^1 - U_-^1\Big) + d_{_{0}} ,\nonumber\\
{\tilde h}_3&=&K\Big[\gamma_{\pm}+  W_+^1 \xi_-^4 -  U_-^1
\eta_+^4+a_{_{0}} \Big(\eta_+^4 -\xi_-^4 \Big)\Big] + \zeta_{_{0}}, \nonumber\\
{\tilde h}_4&=&K\Big(\eta_+^{3} - \xi_-^{3} \Big) + \varrho_{_{0}},\label{d25}
\end{eqnarray}
where $c_{_{0}} $ and $d_{_{0}} $ are even real constants while $\zeta_{_{0}}$ and $\varrho_{_{0}}$ are odd real ones, and the functions
$Q_{\pm}$ and $\gamma_{\pm}$ solve
\begin{eqnarray}
\partial _+ Q_{\pm} & =& W_+^1 {W_+^2}' -  {W_+^1}' W_+^2,  \nonumber\\
\partial _- Q_{\pm} & =& {U_-^1}' U_-^2-U_-^1 {U_-^2}',    \nonumber\\
\partial _+ \gamma_{\pm} & =& {W_+^1}' \eta_+^4-W_+^1 {\eta_+^4}', \nonumber\\
\partial _- \gamma_{\pm} & =& {U_-^1} {\xi_-^4}'-{U_-^1}'  {\xi_-^4},   \label{d26}
\end{eqnarray}
in which primes denote differentiation with respect to the arguments.

At the end of this section, we get the solution of the equations of motion for the dual sigma model in the curved background
(\ref{B12}) by using the duality transformation that follows  two possible decompositions of elements of $l\in D$ as
\begin{eqnarray}
l(\sigma^{+} , \sigma^{-})\; =\; { g}(\sigma^{+} , \sigma^{-}) {\tilde h}(\sigma^{+}
, \sigma^{-})\;=\; {\tilde g}(\sigma^{+} , \sigma^{-}) {h'}(\sigma^{+}
, \sigma^{-}),~~~~~~ {h'}(\sigma^{+}
, \sigma^{-})\in {G}.\label{d27}
\end{eqnarray}
If we write all elements of the Lie supergroups $G$ and $\tilde G$ as a product of elements of one-parametric sub-(super)groups, then, the equation
(\ref{d27}) yields
{\small \begin{eqnarray}
e^{\chi { X}_4}  e^{y { X}_2} e^{x { X}_1}  e^{\psi { X}_3}
e^{{\tilde h}_4 {\tilde X}^4}  e^{{\tilde h}_2 {\tilde X}^2}  e^{{\tilde h}_1 {\tilde X}^1}
e^{{\tilde h}_3 {\tilde X}^3}=e^{\tilde \chi {\tilde X}^4}  e^{\tilde y {\tilde X}^2}  e^{\tilde x {\tilde X}^1}
e^{\tilde \psi {\tilde X}^3} e^{{ h'}^4 {X}_4} e^{{h'}^2 { X}_2}  e^{{ h'}^1 {X}_1} e^{{h'}^3 {X}_3}. \label{d28}
\end{eqnarray}}
Now, using the (anti)commutation relations of the Drinfel'd superdouble $\left(({\cal C}^1_1+{\cal A})\;,\;{\cal I}_{(2|2)}\right)$ obtained in Section 3,
the right hand side of the equation (\ref{d28}) is rewritten in the form
\begin{eqnarray}
e^{{ h'}^4 {X}_4}~ e^{{h'}^2~ { X}_2}  ~e^{{ h'}^1 {X}_1}~ e^{{h'}^3 {X}_3}~&&\hspace{-3mm}
e^{\Big(\tilde \chi + \tilde y e^{{h'}^1} {h'}^3\Big) {\tilde X}^4}  e^{\tilde y e^{{h'}^1} {\tilde X}^2} \nonumber\\
&&\hspace{-7mm}\times e^{\Big(\tilde x-\tilde y  {h'}^2 + \tilde y e^{{h'}^1} {h'}^4 {h'}^3 + e^{{h'}^1}{\tilde \psi } {h'}^3\Big) {\tilde X}^1}
e^{\Big(\tilde y {h'}^4 +  \tilde \psi  \Big)e^{{h'}^1}  {\tilde X}^3}. \label{d29}
\end{eqnarray}
Hence, we get
\begin{eqnarray}
\tilde x &=& {\tilde h}_1 - {\tilde h}_3 \psi + y e^{-x} {\tilde h}_2, \nonumber\\
\tilde y &=&  e^{-x} {\tilde h}_2, \nonumber\\
\tilde \psi &=& e^{-x} ({\tilde h}_3 - {\tilde h}_2 \chi), \nonumber\\
\tilde \chi &=& {\tilde h}_4 - {\tilde h}_2 \psi, \label{d30}
\end{eqnarray}
and
\begin{eqnarray}
{h'}^1 &=&x,~~~{h'}^2\;=\;y,~~ {h'}^3\;=\; \psi,~~~{h'}^4\;=\;\chi. \label{d31}
\end{eqnarray}
Inserting  relations (\ref{d22}) and (\ref{d25}) into (\ref{d30}) we get the solution of the equations of motion for the dual sigma model given by the action
(\ref{B11}) as follows:
\begin{eqnarray}
\tilde x &=& K\Big[Q_{\pm} +   W_+^1 W_+^2-U_-^1 U_-^2 -\frac{(\gamma_{\pm}- W_+^1 \eta_+^{4} +U_-^1 \xi_-^{4})(\eta_+^{3} + \xi_-^{3})}{W_+^1 + U_-^1} \Big],   \nonumber\\
\tilde y &=& K \Big(\frac{W_+^1 - U_-^1}{W_+^1 + U_-^1}\Big), \nonumber\\
\tilde \psi &=&\frac{K}{W_+^1 + U_-^1} \Big(\gamma_{\pm}+   U_-^1\xi_-^{4} - W_+^1\eta_+^{4}\Big), \nonumber\\
\tilde \chi &=& \frac{2K}{W_+^1 + U_-^1} \Big(U_-^1\eta_+^{3}  - W_+^1 \xi_-^{3}\Big), \label{d32}
\end{eqnarray}
where we have assumed that the constants $a_0, b_0, c_0, d_0$ and $\zeta_{_{0}}, \varrho_{_{0}}$ are zero.

Applying the formula (\ref{A4}) for the dual model, the equations of motion of the sigma model given by the curved background
(\ref{B12}) are obtained to be of the form
\begin{eqnarray}
\partial _+ \partial _- \tilde x + (\frac{1}{K+ \tilde y})  \partial _- {\tilde \chi}  \partial _+ {\tilde \psi}+ (\frac{1}{K- \tilde y})
\partial _- {\tilde \psi} \partial _+ {\tilde \chi}&=&0,\nonumber\\
\partial _+ \partial _- \tilde y + 2(\frac{{\tilde y}}{K^2- {\tilde y}^2}) \partial _+ {\tilde y} \partial _- {\tilde y}&=&0,\nonumber\\
\partial _+ \partial _- \tilde \psi + (\frac{1}{K- \tilde y})  \partial _- {\tilde \psi} \partial _+ {\tilde y}-
(\frac{1}{K+ \tilde y}) \partial _- {\tilde y}  \partial _+ {\tilde \psi} &=&0,\nonumber\\
\partial _+ \partial _- \tilde \chi - (\frac{1}{K+ \tilde y})  \partial _- {\tilde \chi} \partial _+ {\tilde y}+
(\frac{1}{K- \tilde y}) \partial _- {\tilde y}  \partial _+ {\tilde \chi} &=&0.\label{d32.1}
\end{eqnarray}
Due to the complexity of the above equations, we are unable to solve them directly. But we have found their solution  in the form (\ref{d32}) by using the flat supercoordinates transformation. One can check that solution (\ref{d32}) actually satisfies the equations of motion (\ref{d32.1}).
\section{Invariance  of the dilaton field  under the supercoordinates  transformation and conformal sigma models}
To guarantee the
conformal invariance of the sigma models, at least at the one--loop level, one must show that they satisfy  the vanishing $\beta$-function equations.
To this end, we need to add a scalar field $\varphi$ (dilaton field) to the action (\ref{A1}) and  rewrite it in the usual standard form \cite{ER5}
\begin{eqnarray}\label{e.0}
S\;=\;\frac{1}{2}\int\!d\sigma^{+} d\sigma^{-}\;\Big[(-1)^{^{\Upsilon}} \partial_{+}\Phi
^{^{\Upsilon}}\;  {
{\cal{E}}}\hspace{-0.5mm}_{_{
\Upsilon \Lambda}}(\Phi)\;\partial_{-}\Phi^{^{\Lambda}}-\frac{1}{4} R^{(2)} \varphi\Big],
\end{eqnarray}
where $R^{(2)}$ is the curvature of the worldsheet. It is well known that the connection between the string effective action and the sigma model (\ref{e.0})
is through the Weyl anomaly coefficients \cite{ER5}
\begin{eqnarray}
\beta_{_{\Upsilon \Lambda}}^{^{(G)}} &:=& R_{_{\Upsilon \Lambda}}+\frac{1}{4} H_{_{\Upsilon \Delta \Xi}}
H^{^{\Xi \Delta }}_{\;\;~~\Lambda}+2\overrightarrow{\nabla}_{_{\Upsilon}}
\overrightarrow{\nabla}_{_{\Lambda}} \varphi=0,\label{e.1}\\
\beta_{_{\Upsilon \Lambda}}^{^{(B)}}  &:=& (-1)^{^{\Delta}}~  \overrightarrow{\nabla}^{^{\Delta}}(
e^{-2\varphi} H_{_{\Delta \Upsilon \Lambda}}) =0,\label{e.2}\\
{{\beta}^{(\varphi)}}  &:=& -R -\frac{1}{12} H_{_{\Upsilon\Lambda\Delta}}H^{^{\Delta\Lambda\Upsilon}}+4 \overrightarrow{\nabla}_{_{\Upsilon}} \varphi \overrightarrow{\nabla}^{^{\Upsilon}} \varphi  -4 \overrightarrow{\nabla}_{_{\Upsilon}} \overrightarrow{\nabla}^{^{\Upsilon}} \varphi=0,\label{e.3}
\end{eqnarray}
where $R_{_{\Upsilon \Lambda}}$ and $R$ are the  Ricci tensor and the scalar curvature, respectively,
and $H_{_{ \Upsilon \Lambda \Delta}}$ is defined by equation (\ref{A3}).  As shown in  \cite{ER5},  the vanishing of $\beta$-function equations can be derived as the equations of motion of the string effective action on the supermanifold $M$
\begin{eqnarray}
S_{_{eff}}~=~ \int d^{(m , n)}\Phi\; \sqrt{G}e^{-2\varphi} \Big(R+4{\overrightarrow{\nabla}_{_{\Upsilon}}} {\varphi} \overrightarrow{\nabla}^{^{\Upsilon}} {\varphi}
+\frac{1}{12} H_{_{\Upsilon\Lambda\Delta}}H^{^{\Delta\Lambda\Upsilon}}\Big),\label{e.4}
\end{eqnarray}
where $(m , n)=(d_B|d_F)$ and $G=sdet ({_{_{\Upsilon}}G_{_{\Lambda}}})$.
As shown in Section 5, the action (\ref{d10}) has been expressed in terms of the flat supercoordinates $\Omega^{^{A}}$  transforming
the flat metric $G_{_{\Upsilon\Lambda}}(\Phi)$ into a constant form $G_{_{AB}}(\Omega)$. Moreover, the model is torsionless. Therefore, ${\bar{\Gamma}}^A_{\;B C}$ and ${\bar{H}}_{_{{A B C}}}$ are vanished. Consequently, equation (\ref{e.2}) is satisfied and equations  (\ref{e.1}) and  (\ref{e.3}) are, respectively, read
\begin{eqnarray}
0&=&  \Big(\frac{ \bar{\varphi}\overleftarrow{\partial}}{\partial
\Omega^{^{A}}} \Big)  \frac{ \overleftarrow{\partial}}{\partial
\Omega ^{^{B}} },\label{e.5}\\
0&=& (-1)^{^{B}}~ \frac{ \bar{\varphi}\overleftarrow{\partial}}{\partial
\Omega^{^{A}}} ~\bar{G}^{^{AB}}~  \frac{\bar{\varphi} \overleftarrow{\partial}}{\partial
\Omega ^{^{B}} }- \Big(\frac{ \bar{\varphi}\overleftarrow{\partial}}{\partial
\Omega^{^{A}}} \Big)  \frac{ \overleftarrow{\partial}}{\partial
\Omega ^{^{B}} }~\bar{G}^{^{BA}}.\label{e.6}
\end{eqnarray}
It is clear that for the metric constant (\ref{d9}) and a constant dilaton field, the vanishing $\beta$-function
equations (equations (\ref{e.5}) and (\ref{e.6}))
are satisfied. But, since we know
the flat supercoordinates of the model,  we can easily find the general form of the dilaton field that together with the metric constant (\ref{d9})
satisfy the equations (\ref{e.1})-(\ref{e.3}). From the form of equations (\ref{e.5}) and (\ref{e.6}) it is easy to  see that the general form of their dilaton field is
\begin{eqnarray}
\bar{\varphi}(\Omega) ~=~ \bar{\varphi}_{_{0}}+ \frac{\lambda_1 \lambda_2}{k_{_{0}}}~
\Omega^{^{1}} + k_{_{0}} \Omega^{^{2}}+\lambda_1 \Omega^{^{3}}+\lambda_2 \Omega^{^{4}},\label{e.7}
\end{eqnarray}
where $\bar{\varphi}_{_{0}}$ and $k_{_{0}}$ are even real constants while $\lambda_1$ and $\lambda_2$ are odd real ones. Now, inserting  the transformation
(\ref{d8}) into  (\ref{e.7}), one gets the general form of the dilaton field of the original sigma model (the model described by the metric (\ref{B7})) in such a way that the result is
\begin{eqnarray}
{\varphi}(\Phi) ~=~ {\varphi}_{_{0}}+ \frac{\lambda_1 \lambda_2}{k_{_{0}}}~
e^{x} + k_{_{0}} (y e^{-x} +\psi \chi)+\lambda_1  \psi e^x +\lambda_2 \chi,\label{e.8}
\end{eqnarray}
where ${\varphi}_{_{0}} = \bar{\varphi}_{_{0}}+ \frac{\lambda_1 \lambda_2}{k_{_{0}}}~
a_{_{0}}  + k_{_{0}} b_{_{0}} +\lambda_1  \alpha_{_{0}}  +\lambda_2 \beta_{_{0}} $.

On the other hand, as explained in subsection (\ref{s3.1}), the original model is  flat (its the Ricci tensor and the scalar curvature are zero) and also torsionless. So, for this model, equation (\ref{e.2}) is satisfied and equations (\ref{e.1}) and (\ref{e.3}) are, respectively, turned into
\begin{eqnarray}
0&=&  (\frac{ {\varphi}\overleftarrow{\partial}}{\partial
\Phi^{^{\Upsilon}}} )  \frac{ \overleftarrow{\partial}}{\partial
\Phi^{^{\Lambda}} }- (\frac{ {\varphi}\overleftarrow{\partial}}{\partial
\Phi^{^{\Delta}}})\Gamma^{^{\Delta}}_{_{~\Upsilon\Lambda}},\label{e.9}\\
0&=& (-1)^{^{\Lambda}}~ \frac{ {\varphi}\overleftarrow{\partial}}{\partial
\Phi^{^{\Upsilon}}} ~{G}^{^{\Upsilon\Lambda}}~  \frac{{\varphi} \overleftarrow{\partial}}{\partial
\Phi^{^{\Lambda}} }- \Big[ (\frac{{\varphi}\overleftarrow{\partial}}{\partial
\Phi^{^{\Upsilon}}})  \frac{ \overleftarrow{\partial}}{\partial
\Phi^{^{\Lambda}} }- (\frac{ {\varphi}\overleftarrow{\partial}}{\partial
\Phi^{^{\Delta}}})\Gamma^{^{\Delta}}_{_{~\Upsilon\Lambda}}\Big] ~{G}^{^{\Lambda\Upsilon}}.\label{e.10}
\end{eqnarray}
It is interesting to see that the dilaton field (\ref{e.8}) is the general solution of equations (\ref{e.9}) and (\ref{e.10}); that is,
the metric (\ref{B7}) together with the dilaton field (\ref{e.8}) satisfy the vanishing $\beta$-function equations. Thus, we conclude that the models (\ref{B6})
and (\ref{d10})  are conformally invariant up
to one-loop.  We also showed that under the transformation of supercoordinates (\ref{d8}), one goes from
${\varphi}(\Phi)$ to $\bar{\varphi}(\Omega)$ and vice versa, i.e., the dilaton field is invariant under the transformation (\ref{d8}).

Let us now turn into the dual model. For the dual model with the background (\ref{B12}), we find that the
only non-zero component of ${\tilde R}_{_{\Upsilon \Lambda}}$
is ${\tilde R}_{22} = \frac{2 K^2}{(K^2-{\tilde y}^2)^2}$; as ${\tilde G}^{22}=0$, ${\tilde R}=0$. Also,  one quickly finds that the only non-zero component of ${\tilde H}$ is ${\tilde H}_{234} =  -\frac{ K^2}{(K^2-{\tilde y}^2)^2}$. It is then straightforward  to verify that
${\tilde H}_{_{\Upsilon\Lambda\Delta}} {\tilde H}^{^{\Delta\Lambda\Upsilon}}=0$, and
the only non-zero component of ${\tilde H}_{_{\Upsilon \Delta \Xi}}
{\tilde H}^{^{\Xi \Delta}}_{_{\;\;~\Lambda}}$ is ${\tilde H}_{234} {\tilde H}^{43}_{\;\;~2}  =  -\frac{ K^2}{(K^2-{\tilde y}^2)^2}$.
Putting these in equations (\ref{e.1})-(\ref{e.3}) we arrive at\footnote{Note that the right covariant derivative of the components of
a three-form  on a supermanifold $M$ is given by
$$
{\tilde H}_{_{\Delta \Upsilon\Lambda}} \overleftarrow{\nabla}_{_{\Xi}}={\tilde H}_{_{\Delta \Upsilon\Lambda}}  \frac{  \overleftarrow{\partial}}{\partial
{\tilde \Phi}^{^{\Xi}}}-(-1)^{^{(\Upsilon+\Lambda)(\Delta+\Omega)}}~{\tilde H}_{_{ \Omega \Upsilon\Lambda}} {\tilde \Gamma}^{^{\Omega}}_{_{~\Delta\Xi}}
-(-1)^{^{\Lambda(\Upsilon+\Omega)}}~{\tilde H}_{_{ \Delta \Omega \Lambda}} {\tilde \Gamma}^{^{\Omega}}_{_{~\Upsilon\Xi}}
-{\tilde H}_{_{ \Delta \Upsilon \Omega }} {\tilde \Gamma}^{^{\Omega}}_{_{~\Lambda \Xi}}.$$}
\begin{eqnarray}
0&=& {\tilde R}_{_{\Upsilon \Lambda}}+\frac{1}{2}{\tilde H}_{_{\Upsilon 34}}
{\tilde H}^{^{43}}_{_{\;\;~\Lambda}}+2\Big[
(\frac{ {\tilde \varphi}\overleftarrow{\partial}}{\partial
{\tilde \Phi}^{^{\Upsilon}}} )  \frac{ \overleftarrow{\partial}}{\partial
{\tilde \Phi}^{^{\Lambda}} }- (\frac{ {\tilde \varphi}\overleftarrow{\partial}}{\partial
{\tilde \Phi}^{^{\Delta}}}) {\tilde \Gamma}^{^{\Delta}}_{_{~\Upsilon\Lambda}}\Big],\label{e.11}\\
0&=& -2(-1)^{^{\Delta}}~(\frac{ {\tilde \varphi}\overleftarrow{\partial}}{\partial
{\tilde \Phi}^{^{\Xi}}} )~ {\tilde G}^{^{\Xi \Delta}} {\tilde H}_{_{ \Delta \Upsilon\Lambda}}+(-1)^{^{\Delta + \Xi+\Xi(\Upsilon+\Lambda)}}
{\tilde G}^{^{\Xi \Delta}} ({\tilde H}_{_{ \Delta \Upsilon\Lambda}} \overleftarrow{\nabla}_{_{\Xi}}),
\label{e.12}\\
0&=& (-1)^{^{\Lambda}}~ \frac{ {\tilde \varphi}\overleftarrow{\partial}}{\partial
{\tilde \Phi}^{^{\Upsilon}}} ~{\tilde G}^{^{\Upsilon\Lambda}}~  \frac{{\tilde \varphi} \overleftarrow{\partial}}{\partial
{\tilde \Phi}^{^{\Lambda}} }- \Big[ (\frac{{\tilde \varphi}\overleftarrow{\partial}}{\partial
{\tilde \Phi}^{^{\Upsilon}}})  \frac{ \overleftarrow{\partial}}{\partial
{\tilde \Phi}^{^{\Lambda}} }- (\frac{ {\tilde \varphi}\overleftarrow{\partial}}{\partial
{\tilde \Phi}^{^{\Delta}}}) {\tilde \Gamma}^{^{\Delta}}_{_{~\Upsilon\Lambda}}\Big] ~{\tilde G}^{^{\Lambda\Upsilon}}.\label{e.13}
\end{eqnarray}
By solving the above equations, one gets the general form of the dilaton field of the dual model to be of the form
\begin{eqnarray}
{\tilde \Phi}~=~{\tilde \Phi_0} + \frac{{\tilde k}_{_{0}}}{2K} \ln \Big|\frac{K+\tilde y}{K-\tilde y}\Big|
-\frac{3}{32} \Big(\ln \Big|\frac{K+\tilde y}{K-\tilde y}\Big|\Big)^2,\label{e.14}
\end{eqnarray}
where ${\tilde \Phi}_{_{0}}$ and ${\tilde k}_{_{0}}$ are the even constants of integration.
Thus, conformal invariance of the dual model is also guaranteed up to one-loop.

\vspace{2mm}
\section{Conclusion}

In the present work, we have obtained the solution  of the equations of motion for a super non-Abelian T-dual  sigma model in curved background by
the super Poisson-Lie T-duality.
In this way, we have first used the fact
that the super Poisson-Lie T-duality transformations can be explicitly performed for the Drinfel'd superdouble
$\left((C^1_1+A)~,~I_{(2|2)}\right)$. Then,
we have obtained the explicit transformation between the supergroup
coordinates of the model living in the flat background and its
flat supercoordinates. Also, we have proved that flat supercoordinates transformation is a supercanonical transformation.
Furthermore, we have found a linear transformation of the flat supercoordinates which reduces
the action of the flat model with the metric constant to the Polyakov action on a  $(2|2)$-dimensional flat supermanifold.
By reducing the transformation of supercoordinates to the solution of the wave equation and
using the super Poisson-Lie T-duality transformations, we have
obtained the solution of the equations of motion of the dual sigma model. Finally,
we were able to find an example of the super non-Abelian T-dual conformal sigma models
(at least at the one-loop level) for which the vanishing $\beta$-function equations are satisfied.
By solving the vanishing $\beta$-function equations we found the general form of the dilaton field of the models and showed that
the dilaton field of the original model is invariant under the flat supercoordinates transformation; that is, one can  go from
${\varphi}(\Phi)$ to $\bar{\varphi}(\Omega)$ and vice versa.

\acknowledgments

The author would like to thank the anonymous referee for invaluable comments and criticisms.
Also, the author is grateful to A. Rezaei-Aghdam for his valuable comments
and H. Shams for carefully reading the
manuscript.



\begin{thebibliography}{99}

\bibitem{Henneaun}
M. Henneaux, L. Mezincescu, \emph{A $\sigma$-model interpretation of Green-Schwarz covariant
superstring action}, \emph{Phys. Lett. B} {\bf 152} (1985) 340-342.

\bibitem{Read} N. Read and H. Saleur,
\emph{Exact spectra of conformal supersymmetric nonlinear sigma models in two dimensions},
\emph{Nucl. Phys. B} {\bf 613} (2001) {409-444}  [hep-th/0106124]; S. Guruswamy,
A. LeClair and A. W. W. Ludwig,
\emph{$gl(N|N)$ Super-current algebras for disordered Dirac fermions in two dimensions},
\emph{Nucl. Phys. B} {\bf 583} (2000) {475-512}  [cond-mat/9909143].


\bibitem {Tseytlin} R. R. Metsaev  and  A. A. Tseytlin, \emph{Type IIB superstring action in $AdS_5 \times S^5$ background},
\emph{Nucl. Phys. B} {\bf 533} (1998) {109} [hep-th/9805028].

\bibitem {Berkovits}  N. Berkovits, C. Vafa and E. Witten,
\emph{Conformal field theory of AdS background with Ramond-Ramond flux}, \emph{JHEP} {\bf 03} (1999) 018  [hep-th/9902098].

\bibitem{Bershadsky} M. Bershadsky,  S. Zhukov and A. Vaintrob,
\emph{$PSL(n|n)$ sigma model as a conformal field theory},
\emph{Nucl. Phys. B} {\bf 559} (1999) {205-234} [hep-th/9902180].


\bibitem{Zwiebach} N. Berkovits, M. Bershadsky, T. Hauer, S. Zhukov and B. Zwiebach,
\emph{Superstring theory on
$AdS_2 \times S^2$ as a coset supermanifold},
\emph{Nucl. Phys. B} {\bf 567} (2000) {61-86} [hep-th/9907200].

\bibitem{Boer} J. de Boer and S. L. Shatashvili,
\emph{Two-dimensional conformal field theories on $AdS(2d + 1)$ backgrounds},
\emph{JHEP} {\bf 06} (1999)  013 [hep-th/9905032].

\bibitem{Kagan} David Kagan and Charles A. S. Young,
\emph{Conformal sigma models on supercoset targets},
\emph{Nucl. Phys. B} {\bf 745} (2006) {109-122} [hep-th/0512250].


\bibitem{Buscher} T. H. Buscher,  \emph{A symmetry of the string background field equations}, \emph{Phys. Lett.
B} {\bf 194} (1987) 59; \emph{Path integral derivation of quantum duality in nonlinear $\sigma$-models}, \emph{Phys. Lett. B} {\bf 201} (1988) 466.

\bibitem{Duff} M.J. Duff, \emph{Duality rotations in string theory}, \emph{Nucl. Phys. B} {\bf 335} (1990) 610-621.


\bibitem{Giveon} A. Giveon and M. Ro\v{c}ek, \emph{Generalized duality in curved string backgrounds}, 
\emph{Nucl. Phys. B} {\bf 380} (1992) 128 [hep-th/9112070].

\bibitem{Kiritsis} E. Kiritsis, \emph{Exact duality symmetries in CFT and string theory},
\emph{Nucl. Phys. B} {\bf 405} (1993) 109 [hep-th/9302033].


\bibitem{Rocek} M. Ro\v{c}ek and E. Verlinde, \emph{Duality, quotients and currents},
\emph{Nucl. Phys. B} {\bf 373} (1992) 630  [hep-th/9110053].

\bibitem{Ossa} X.C. de la Ossa and F. Quevedo, \emph{Duality symmetries from non-abelian isometries in string
theory}, \emph{Nucl. Phys. B} {\bf 403} (1993) 377 [hep-th/9210021].

\bibitem{klse:dna}
C.~Klim\v{c}\'{\i}k and P.~\v{S}evera, \emph{Dual non--{A}belian duality and the {D}rinfeld double}, \emph{Phys.
Lett. B} { \bf 351} (1995) 455 [hep-th/9502122].

\bibitem{Drinfel'd}
V. G. Drinfeld, \emph{Quantum groups}, in the Proceedings of the International Congress of
Mathematicians, Berkeley U.S.A. (1986), American Mathematical Society, New York U.S.A. (1987), pg. 798.



\bibitem{hla:slnbytduality} L. Hlavat\'y, \emph{Classical solution of a sigma model in curved background}, \emph{Phys.
Lett. B} {\bf 625} (2005) 285 [hep-th/0506188].

\bibitem{hla1} L. Hlavaty, J. Hybl and  M. Turek, \emph{Classical solutions of sigma models in curved backgrounds by the Poisson-Lie T-plurality}, \emph{Int. J.
Mod. Phys. A} {\bf 22} (2007) 1039 [hep-th/0608069].

\bibitem{hlasno:3dsm2} L. Hlavat\'y and L. \v Snobl, \emph{Poisson--Lie T--plurality of
three--dimensional conformally invariant sigma models II: Nondiagonal metrics and dilaton puzzle}, \emph{JHEP} {\bf 10 } (2004) 045 [hep-th/0408126].

\bibitem{hla2} L. Hlavat\'y and M. Turek, \emph{Flat coordinates and dilaton fields for three-dimensional conformal sigma models}, \emph{JHEP} {\bf 06 } (2006)
003; L. Hlavat\'y, \emph{Dilatons in curved backgrounds by the Poisson-Lie transformation},  [hep-th/0601172].

\bibitem {ER}  A. Eghbali and A. Rezaei-Aghdam, \emph{Poisson-Lie T-dual sigma models on supermanifolds},
\emph{JHEP} { \bf 09} (2009) {094 }  [arXiv:0901.1592].

\bibitem {ER5}  A. Eghbali and A. Rezaei-Aghdam, \emph{String cosmology from Poisson-Lie T-dual sigma models on supermanifolds},
\emph{JHEP} {\bf 01} (2012) {151}  [arXiv:1107.2041].

\bibitem {ER8}  A. Eghbali and A. Rezaei-Aghdam, \emph{WZW models as mutual super Poisson-Lie T-dual sigma models},
\emph{JHEP} {\bf 07} (2013) {134}  [arXiv:1303.4069].

\bibitem {ER7}  A. Eghbali and A. Rezaei-Aghdam, \emph{Super Poisson-Lie symmetry of the $GL(1|1)$ WZNW model and worldsheet boundary conditions},
\emph{Nucl. Phys. B} {\bf 866} (2013)  26-42  [arXiv:1207.2304].

\bibitem{B}  N. Backhouse, \emph{A classification of four-dimensional Lie
superalgebras},  \emph{J. Math. Phys.}  {\bf 19} (1978)
2400-2402.

\bibitem{D}  B. DeWitt, \emph{Supermanifolds}, Cambridge University Press 1992.

\bibitem {ER4} A. Eghbali, A. Rezaei-Aghdam and F. Heidarpour,  \emph{Classification of four and six dimensional Drinfel'd
superdoubles}, \emph{J. Math. Phys.} {\bf 51} (2010) 103503
[arXiv:0911.1760].

\bibitem {ER1} A. Eghbali, A. Rezaei-Aghdam and  F. Heidarpour,
\emph{Classification of two and three dimensional Lie
super-bialgebras},   \emph{J. Math. Phys.} {\bf 51} (2010) 073503
[arXiv:0901.4471].

\bibitem{klim1}
C.~Klim\v{c}\'{\i}k and P.~\v{S}evera, \emph{Poisson-Lie T-duality and loop groups of Drinfeld doubles}, \emph{Phys.
Lett. B} {\bf 372} (1996) 65-71 [hep-th/9512040].

\bibitem{klim2}
C.~Klim\v{c}\'{\i}k and P.~\v{S}evera, \emph{Poisson-Lie T-duality: Open strings and D-branes}, \emph{Phys.
Lett. B} { \bf 376} (1996) 82-89  [hep-th/9512124].


\bibitem{Tokunaga}
 T. Tokunaga, \emph{String Theories on Flat Supermanifolds}, Report Number YITP-05-47, [hep-th/0509198].




\end{thebibliography}
\end{document}